\documentclass[%
 reprint,
 amsmath,amssymb,
 aps,
]{revtex4-2}

\usepackage{graphicx}% Include figure files
\usepackage{dcolumn}% Align table columns on decimal point
\usepackage{bm}% bold math
\usepackage{xcolor}
\usepackage{dsfont}
\usepackage[utf8]{inputenc}
\usepackage[T1]{fontenc}
\usepackage[english,serbian]{babel}
\usepackage[OT1]{fontenc}
\usepackage{mathrsfs}
\usepackage{graphicx}
\usepackage{amssymb}
\usepackage{amsbsy}
\usepackage{latexsym}
\usepackage{mathtools}
\usepackage{titlesec}
\usepackage{amsmath}
\usepackage{color}
\usepackage[utf8]{inputenc}
\usepackage{lipsum}
\usepackage{fontenc}
\usepackage{bbold}

\usepackage{chngcntr}
\numberwithin{equation}{section}
\renewcommand{\theequation}{\arabic{section}.\arabic{equation}}
\def\del{\partial}

\def\dj{d\kern-0.4em\char"16\kern-0.1em}

\def\diff{\textrm{d}}
\def\Diff{\textrm{D}}

\def\tr{{\rm Tr}}

\def\dj{d\kern-0.4em\char"16\kern-0.1em}
\def \Dj {\mbox{\raise0.3ex\hbox{-}\kern-0.4em D}}

\begin{document}

\preprint{APS/123-QED}

\title{Noncommutative $D=5$ Chern-Simons Gravity: Kaluza-Klein Reduction and Chiral Gravitational Anomaly}

\author{Du\v{s}an \Dj or\dj evi\'{c} and Dragoljub Go\v{c}anin}
\affiliation{Faculty of Physics, University of Belgrade, Studentski Trg 12-16, 11000 Belgrade, Serbia}

\email{dgocanin@ipb.ac.rs}

\selectlanguage{english}

%\date{\today}

\begin{abstract}

Actions for noncommutative (NC) gauge field theories can be expanded perturbatively in powers of the noncommutativity parameter $\theta$ using the Seiberg-Witten map between ordinary classical fields and their NC counterparts. The leading order term represents classical ($\theta=0$) action while higher-order terms give us $\theta$-dependent NC corrections that ought to capture some aspects of quantum gravity. Building on previous work of Aschieri and Castellani on NC Chern-Simons (CS) gauge and gravity theories, showing that non-trivial $\theta$-dependence exists only for spacetime dimensions $D\geq 5$, we investigate a correlated effect of these extra spatial dimensions and noncommutativity on four-dimensional physics, up to first-order in $\theta$. Assuming that one spatial dimension is compactified into a circle, we apply the Kaluza-Klein reduction procedure on the NC $D=5$ CS theory for the conformal gauge group $SO(4,2)$, to obtain an effective, $\theta$-dependent four-dimensional theory of gravity that has Einstein-Hilbert gravity with negative cosmological constant as its commutative limit. We derive field equations for this modified theory of gravity and study the effect of NC interactions on some classical geometries, such as the AdS-Schwarzschild black hole. We find that this NC background spacetime gives rise to chiral gravitational anomaly due to the nonvanishing $\theta$-dependent Pontryagin density.     

\end{abstract}

\maketitle

\selectlanguage{english}

\section{Introduction}

Higher-dimensional theories of gravity have been of interest in physics ever since Kaluza and Klein proposed an elegant way of unifying Einstein's theory of General Relativity with Maxwell's electromagnetism, in terms of a purely geometric theory of five-dimensional gravity \cite{Kaluza, Klein} with one spatial dimension compactified into a circle \textemdash the method now known as the Kaluza-Klein (KK) dimensional reduction.  
In search for a Grand Unified Theory with appropriate four-dimensional phenomenology, extra spatial dimensions became a topic of extensive investigation that has led to a plethora of higher-dimensional supergravity models \cite{SUGRA}. Some of these models turned out to be low-energy approximations to certain types of superstring theory, where extra spatial dimensions most commonly and unavoidably arise. For a comprehensive account on the development of the idea of extra spatial dimensions and KK reduction see \cite{KK}.    

The most natural higher dimensional ($D>4$) generalization of the Einstein-Hilbert (EH) Lagrangian (with the cosmological constant term)
that does not involve torsion and gives at most second-order field equations for the metric tensor in the torsion-free sector, is the Lovelock-Lanczos (LL) Lagrangian \cite{Lovelock, Lanczos, CS_book}, 
\begin{equation}\label{LLL}
L^{(D)}_{LL}=\sum\limits_{i=0}^{[D/2]}c_{i}L_{i}^{(D)},
\end{equation}
with locally Lorentz-invariant Lagrangian $D$-forms 
\begin{equation}
L_{i}^{(D)}=\varepsilon_{a_{1}\dots a_{D}}R^{a_{1}a_{2}}\dots R^{a_{2i-1}a_{2i}}e^{a_{2i+1}}\dots e^{a_{D}}.    
\end{equation}
It is expressed in a coordinate-free fashion of the first-order formalism (exterior product between forms is implied) where vielbein (tetrade in $D=4$) $1$-form $e^{a}=e^{a}_{\mu}\diff x^{\mu}$ and spin-connection $1$-form $\omega^{ab}=\omega^{ab}_{\mu}\diff x^{\mu}$ are treated as independent dynamical fields. This renders the LL action manifestly invariant under diffeomorphisms. The curvature $2$-form is $R^{ab}=\diff\omega^{ab}+\omega^{a}_{\;\;c}\omega^{cb}$, and parameters $c_{i}$ are arbitrary constants with dimensions $[c_{i}]=[length]^{2i-D}$ (since by dimensional analysis we have $[e]=[length]^{1}$ and $[\omega]=[length]^{0}$). 

There is a long tradition of attempts to formulate gravity as a proper gauge field theory \cite{Blagojevic}. A well-known example of a gauge field theory that can only be defined on odd-dimensional spacetimes, i.e. for $D=2n-1$, is the Chern-Simons (CS) theory. The CS Lagrangian
can be regarded as a particular case of the LL Lagrangian (\ref{LLL}) with a unique choice of parameters that renders CS theory invariant under an enlarged gauge group of transformations (one that contains the Lorentz group $SO(D-1,1)$ as a subgroup). This enhancement follows from the fact that, in CS theory, the vielbein and the spin-connection stand on equal footing, as components of the connection of an enlarged gauge (super)-group, which makes CS theory a proper gauge theory of (super)-gravity \cite{CS, CS_SUGRA1, CS_SUGRA2, CS_SUGRA3}, at least in odd dimensions.  
All parameters in the CS Lagrangian are fixed rational coefficients that cannot be changed without breaking the gauge invariance. Moreover, the fact that gauge invariance is satisfied off-shell makes CS theory a promising candidate for a quantum theory of gravity in odd-dimensional spacetimes. However, full quantization of CS theory in dimensions greater than three is not well-understood \cite{CS_book, Witten:1988hc, Cham1, One-loop}. 

CS Lagrangians describe topological gravity in any odd number of dimensions, suggesting that CS gravity cannot be used to describe $D=4$ phenomenology but only as a toy model for better understanding gravity in general. However, in \cite{Cham2} Chamseddine proposed a model of topological gravity for all even dimensions, which involves a scalar field, and can be obtained by KK reduction from a higher-dimensional CS action (see also \cite{Morales}). This relation explains the appearance of a scalar field and indicates that CS theory can be of even greater importance than previously anticipated. In particular, starting from a $D=5$ CS action for the conformal gauge group $SO(4,2)$ (isometry group of anti-De Sitter (AdS) space $AdS_{5}$), 
the KK reduction procedure yields a $D=4$ topological gravity action with local $SO(3,2)$ gauge symmetry.
While it is possible to consider this action on its own, we encounter only $SO(3,1)$ local Lorentz symmetry when describing gravitational physics on common energy scales, suggesting that $SO(3,2)$ should be broken in some way. Theories based on broken $SO(3,2)$ gauge symmetry have been considered in the literature for a long time, starting with the work of MacDowell and Mansouri \cite{MacDowell-Mansouri} and Stelle and West \cite{stelle-west}, see also \cite{Wilczek, Mukhanov1, Mukhanov2}.
By fixing the value of the scalar field in the topological action proposed in \cite{Cham2}, one obtains the standard EH term with a negative cosmological constant and a topological Gauss-Bonnet term that does not affect classical field equations. 
KK reduction of higher-dimensional topological actions is also studied in \cite{Wu1, Wu2}. 
In this paper, we apply the KK reduction procedure on an NC extension of $D=5$ CS action to see the effects of spacetime noncommutativity in the low-energy, four-dimensional theory of gravity.

It is generally believed that the classical description of spacetime as a smooth manifold breaks at small enough length scales. One way to formalize a deviation from classical geometry is to introduce an abstract algebra of noncommuting coordinates $\hat{x}^{\mu}$, describing a noncommutative (NC) spacetime. These NC coordinates satisfy some non-trivial commutation relations, the simplest of which are the canonical, or $\theta$-constant commutation relations,
\begin{equation}\label{NC_relations}
[\hat{x}^{\mu},\hat{x}^{\nu}]=i\theta^{\mu\nu},  \end{equation}
where NC deformation parameters $\theta^{\mu\nu}$ comprise a constant antisymmetric matrix. The NC parameters are of order $l_{NC}^{2}$, where $l_{NC}$ is an undetermined length scale associated with noncommutativity. In a NC spacetime, individual points cannot be sharply defined, as encoded by a kind of uncertainty relations,  $\Delta\hat{x}^{\mu}\Delta\hat{x}^{\nu}\geq\frac{1}{2}\vert\theta^{\mu\nu}\vert$.
A way to implement this canonical spacetime noncommutativity is to keep the commutative (i.e. classical) structure of spacetime and instead deform the algebra of functions of commuting coordinates, $x^{\mu}$, by substituting the ordinary pointwise multiplication with the noncommutative, but associative, Moyal-Weyl-Groenewold (MWG) $\star$-product,   
\begin{align}\label{Moyal}
(f\star g)(x)&=f(x)e^{\frac{i}{2}\overleftarrow{\partial}_{\mu}\theta^{\mu\nu}\overrightarrow{\partial}_{\nu}}g(x)
\\
&=f(x)g(x)+\frac{i}{2}\theta^{\mu\nu}\partial_{\mu}f(x)\partial_{\nu}g(x)+\dots.\nonumber
\end{align}
Applied on coordinate functions, it gives us the $\theta$-constant $\star$-commutator relations,
\begin{equation}\label{star_commutator}
[x^{\mu},x^{\nu}]_{\star}=x^{\mu}\star x^{\nu}-x^{\nu}\star x^{\mu}=i\theta^{\mu\nu}.   
\end{equation}
For a comprehensive review of the subject, see \cite{NC_book, Leo_NC}. By definition, the MWG $\star$-product is associated with a particular system of coordinates, depending on the choice of noncommuting coordinates $\hat{x}^{\mu}$ on which we impose the $\theta$-constant noncommutativity (\ref{NC_relations}). However, as we will discuss later, the $\star$-product structure can be generalized to any coordinate system in a way that allows us to study NC gauge field theory and NC (super)-gravity in a coordinate-free formulation; see, for example, \cite{Aschieri:2011ng}. In a nongravitational setting, NC three-dimensional Chern-Simons theory was also considered in  \cite{Sheikh-Jabbari:2001nlh}. Also, the subject of NC extra dimensions is treated in \cite{NC_extra1}.  

NC gravity has been studied extensively for the past twenty years and from various viewpoints. In \cite{SWmapApproach1, SWmapApproach2, SWmapApproach3} an approach based on the Seiberg-Witten map was applied to deform pure Einstein gravity. The twist approach was applied in \cite{TwistApproach1, TwistApproach2, TwistSolutions1, TwistSolutions2}. Some other interesting proposals can be found in \cite{Others1, Others2, Others3, Others4, Others5, Others6, Others7}.
NC gravity based on canonically deformed Lorentz symmetry is developed in \cite{PLM-13}, and for canonically deformed AdS group $SO(3,2)$ in \cite{MiAdSGrav, MDVR-14, UsLetter, Us-16}, both models predicting that first non-vanishing NC correction appears
at second-order in $\theta$.
Finally, the NC extension of $D=5$ CS gravity for the conformal gauge group $SO(4,2)$ is constructed by Aschieri and Castellani in \cite{Leo}, and the first-order NC correction is computed explicitly. Their result will be taken as the starting point of this paper. Assuming that one spatial dimension is compactified into a circle, we will derive an effective four-dimensional theory of gravity that amounts to a modification of Einstein's gravity by $\theta$-dependent perturbative NC corrections that can be interpreted as new gravitational interaction terms, capturing some aspects of quantum gravity. We will derive modified field equations for tetrade and spin-connection from this NC action and analyze some of their solutions. We find that NC-modified AdS-Schwarzschild geometry exhibits $\theta$-dependent chiral gravitational anomaly that makes an interesting phenomenological aspect of the theory.    

The paper is organized as follows. In the following section, we briefly review the classical CS theory and, in particular, the $D=5$ case and its KK reduction. In Section $3$, we go through some basic elements of the NC $\star$-product formalism and geometric Seiberg-Witten map, closely following the account of \cite{Leo}. Section $4$ contains the KK reduction of the NC $D=5$ CS action and derivation of the NC-modified field equations for the effective four-dimensional theory of gravity. In Section $5$, we study the NC-correction to AdS-Schwarzschild geometry and compute the $\theta$-dependent Pontryagin topological invariant that gives rise to a chiral gravitational anomaly. Finally, we discuss some main points of the paper and possibilities for further research in Section $6$. Some useful formulae and calculations are given in Appendices A-C. 

%%%%%%%%%%%%%%%%%%%%%%%%%%%%%

\section{Classical CS gauge field theory in $D=5$ and its Kaluza-Klein reduction}

In doing classical (i.e. commutative) gauge field theory, one usually starts with a Lie algebra $\mathfrak{g}$ of some gauge group $\mathcal{G}$, having a set of generators $\{T_{K}\}$ (we take them to be anti-hermitian). Assuming that $\mathcal{G}$-bundle is trivial, there is a globally defined $\mathfrak{g}$-valued gauge field $1$-form, $A=A^{K}T_{K}$, and the corresponding $\mathfrak{g}$-valued field strength $2$-form, $F=\diff A+A\wedge A=F^{K}T_{K}$. Under an infinitesimal gauge transformations, with a $\mathfrak{g}$-valued gauge parameter $0$-form $\epsilon=\epsilon^{K}T_{K}$, they change as 
\begin{align}\label{varA}
    \delta_{\epsilon}A&=-\diff\epsilon-[A,\epsilon]=-\diff\epsilon-A\wedge\epsilon+\epsilon\wedge A,\\
    \delta_{\epsilon}F&= [\epsilon,F]=\epsilon\wedge F-F\wedge\epsilon,\label{varF}
\end{align}
and the algebra of infinitesimal gauge transformations closes in the Lie algebra, 
\begin{equation}
[\delta_{\epsilon_{1}},\delta_{\epsilon_{2}}]=\delta_{-[\epsilon_{1},\epsilon_{2}]}.    
\end{equation}
Covariant derivative acts on $F$ (or any field in the adjoint representation) as \begin{equation}
\Diff F=\diff F+[A,F]=\diff F+A\wedge F-F\wedge A.
\end{equation}
One can define a gauge-invariant topological action over some $2n$-dimensional manifold $\mathcal{M}_{2n}$ with boundary by taking a $2n$-form $\tr\left(F^{n}\right)$ as a Lagrangian, i.e. 
\begin{equation}\label{CS_Tr}
\int\limits_{\mathcal{M}_{2n}}\tr\left(F^{n}\right),     
\end{equation}
where we take $\tr$ in some matrix representation of the Lie algebra (more generally, one can use any symmetric invariant tensor of rank $n$, $\langle\dots\rangle_{n}\colon\mathfrak{g}\times\dots\times\mathfrak{g}\rightarrow\mathbb{C}$, but since we will be using an explicit representation later on, the trace operator is a natural choice \cite{Bertlmann}). Action (\ref{CS_Tr}) is manifestly gauge-invariant, given the gauge transformation law for $F$ and graded cyclicity of the trace. Some basic definitions involving Lie algebra-valued forms are given in Appendix A.

From the Chern-Weil theorem \cite{Bertlmann} follows that  $\tr\left(F^{n}\right)$ can be expressed as an exact form, being an exterior derivative of a Chern-Simons $(2n-1)$-form,    
\begin{equation}\label{basic}
\tr(F^{n})=\diff Q^{(2n-1)}_{CS},    
\end{equation}
where the CS form is given by, 
\begin{equation}
Q^{(2n-1)}_{CS}=n\int\limits_{0}^{1}t^{n-1}\tr\left( A(F+(t-1)A^{2})^{n-1}\right) dt.
\end{equation}
In particular, in five dimensions (for $n=3$), we have
\begin{equation}
Q^{(5)}_{CS}=\tr\left( F^{2}A-\frac{1}{2}FA^{3}+\frac{1}{10}A^{5}\right).
\end{equation}
The topological action (\ref{CS_Tr}) is therefore related by Stokes theorem with the CS action on the $(2n-1)$-dimensional boundary of $\mathcal{M}_{2n}$, 
\begin{equation}
S^{(2n-1)}_{CS}=\alpha\int\limits_{\del{\mathcal{M}_{2n}}}Q^{(2n-1)}_{CS}=\int\limits_{\del{\mathcal{M}_{2n}}}L^{(2n-1)}_{CS},
\end{equation}
where the overall factor $\alpha$ is introduced for later convenience. The gauge invariance of $\tr\left(F^{n}\right)$ implies that the gauge variation of the CS Lagrangian is locally exact, and so 
\begin{equation}
\delta_{\epsilon} S_{CS}^{(2n-1)}=0. \end{equation}
It is worth noting that CS theory need not be regarded as a boundary theory. In particular, it can be defined on an arbitrary $(2n-1)$-dimensional manifold with boundary, in which case (at least for $n=2$) the boundary theory corresponds to the chiral Wess-Zumino-Witten (WZW) model \cite{Elitzur:1989nr}.  
  
Gauge group that we are interested in is the conformal group $SO(4,2)$, i.e. the group of isometries of $AdS_{5}$. Let $J_{\mathbb{A}\mathbb{B}}=(J_{AB},J_{A5})$ be the generators of this group with $SO(4,2)$ group index $\mathbb{A}=(A,5)$ and $A=0,1,2,3,4$, satisfying the $\mathfrak{so}(4,2)$ algebra,
\begin{equation}
[J_{\mathbb{A}\mathbb{B}},J_{\mathbb{C}\mathbb{D}}]=G_{\mathbb{A}\mathbb{D}}J_{\mathbb{B}\mathbb{C}}+G_{\mathbb{B}\mathbb{C}}J_{\mathbb{A}\mathbb{D}}-(\mathbb{C}\leftrightarrow \mathbb{D})
\end{equation}
with $G_{\mathbb{A}\mathbb{B}}=(-++++-)$, or, in a more explicit form,
\begin{align}
&[J_{AB},J_{CD}]=G_{AD}J_{BC}+G_{BC}J_{AD}-(C\leftrightarrow D), \nonumber \\
&[J_{AB},J_{C5}]=G_{BC}J_{A5}-G_{AC}J_{B5}, \nonumber \\
&[J_{A5},J_{C5}]=J_{AC},  
\end{align}
with $G_{AB}=(-++++)$. A representation of this algebra is provided by five-dimensional $4\times 4$ gamma matrices $\Gamma_{A}$, satisfying Cliford algebra $\{\Gamma_{A},\Gamma_{B}\}=2G_{AB}\mathds{1}_{4\times 4}$. In terms of these gamma matrices, the generators are given by $J_{AB}=\frac{1}{2}\Gamma_{AB}=\frac{1}{4}[\Gamma_{A},\Gamma_{B}]$ and $J_{A5}=\frac{1}{2}\Gamma_{A}$, with hermitian relations $J_{AB}^{\dagger}=\Gamma_{0}J_{AB}\Gamma_{0}$ and $J_{A5}^{\dagger}=
\Gamma_{0}J_{A5}\Gamma_{0}$. The five-dimensional gamma-matrices can be defined in terms of familiar, reversed signature, four-dimensional Dirac gamma-matrices as $\Gamma_{A}=(\Gamma_{a}=-i\gamma_{a},\Gamma_{4}=\gamma_{5})$, with $SO(3,1)$ Lorentz index $a=0,1,2,3$. 

The $SO(4,2)$ gauge field can be decomposed as
\begin{equation}
A=\frac{1}{2}\Omega^{AB}J_{AB}+\frac{1}{l}E^{A}J_{A5},    
\end{equation}
where $l$ is the AdS radius. Field strength $2$-form is then
\begin{align}
F&=\diff A+A\wedge A=\frac{1}{2}F^{AB}J_{AB}+F^{A5}J_{A5}\nonumber\\
&=\frac{1}{2}\left(R^{AB}+\frac{1}{l^{2}}E^{A}E^{B}\right)J_{AB}+\frac{1}{l}T^{A}J_{A5},
\end{align}
with curvature and torsion given by
\begin{align}
&R^{AB}=\diff\Omega^{AB}+\Omega^{A}_{\;\;C}\Omega^{CB},\\
&T^{A}=\Diff_{\Omega} E^{A}=\diff E^{A}+\Omega^{A}_{\;\;B}E^{B}.
\end{align}
Using the trace identities (see Appendix B) we compute 
\begin{align}
\tr\left(F^{3}\right)=\frac{3i}{8}\varepsilon_{ABCDE}F^{AB}F^{CD}F^{E5}.
\end{align}
After some partial integration, the CS action in $D=5$ can be cast in the following form, 
\begin{align}\label{CS_first}
S_{CS}^{(5)}=\frac{k}{8}&\int\varepsilon_{ABCDE}\Bigg(\frac{1}{l}R^{AB}R^{CD}E^{E}\\
+&\frac{2}{3l^{3}}R^{AB}E^{C}E^{D}E^{E}+\frac{1}{5l^{5}}E^{A}E^{B}E^{C}E^{D}E^{E}\Bigg), \nonumber
\end{align}
where we took $\alpha=-ik/3$, the dimensionless parameter $k$ being the CS level. In a perhaps more familiar second-order formulation, the action is
\begin{align}\label{CS_second}
S_{CS}^{(5)}=\frac{1}{16\pi G^{(5)}}&\int\diff^{5}x\sqrt{-g}\Big[R-2\Lambda\\
+&\frac{l^{2}}{4}\left(R^{2}-4R^{\mu\nu}R_{\nu\mu}+R^{\mu\nu\rho\sigma}R_{\rho\sigma\mu\nu}\right)\Big], \nonumber \end{align}
with five-dimensional gravitational constant $G^{(5)}=\frac{l^{3}}{8\pi k}$ and cosmological constant $\Lambda=-3/l^{2}$. The last term in (\ref{CS_second}) is the Gauss-Bonnet term, which in $D=5$ is not topological. Note, however, that the action (\ref{CS_second}) is defined on a Riemann-Cartan spacetime, as the torsion is not set to zero, even on-shell. This implies that tensor $R_{\mu\nu\rho\sigma}$ is a Riemman-Cartan curvature tensor, and does not have all symmetry properties of a Riemman curvature tensor on a Riemannian (Lorentzian) spacetime.

Now we consider in some detail the KK dimensional reduction of the $D=5$ CS action (\ref{CS_first}) along the lines of Chamseddine's paper \cite{Cham2}. First, we separate the five-dimensional spacetime coordinates as $x^{\widetilde{\mu}}=(x^{\mu},x^{4})$ where $\mu=0,1,2,3$. Assuming that the one extra spatial dimension is compactified into a circle of radius $R$, i.e. $x^{4}\sim x^{4}+2\pi R$, and taking only KK zero-modes into account (nothing depends on $x^{4}$), we can integrate out the $x^{4}$ coordinate to get an overall factor of $2\pi R$.  

The $SO(4,2)$ gauge field $A=\tfrac{1}{2}\Omega^{AB}J_{AB}+l^{-1}E^{A}J_{A5}$ with $A=(a,4)$ and $a=0,1,2,3$, can be decomposed as
\begin{align}
A=&\frac{1}{2}\Omega_{\mu}^{ab}J_{ab}\diff x^{\mu}+\Omega_{\mu}^{a4}J_{a4}\diff x^{\mu}    \nonumber\\
+&\frac{1}{2}\Omega_{4}^{ab}J_{ab}\diff x^{4}+\Omega_{4}^{a4}J_{a4}\diff x^{4} \nonumber\\
+&\frac{1}{l}E_{\mu}^{a}J_{a5}\diff x^{\mu}+\frac{1}{l}E_{\mu}^{4}J_{45}\diff x^{\mu} \nonumber\\
+&\frac{1}{l}E_{4}^{a}J_{a5}\diff x^{4}+\frac{1}{l}E_{4}^{4}J_{45}\diff x^{4}.
\end{align}
Components $\Omega_{\mu}^{ab}\equiv \omega_{\mu}^{ab}$ and $E_{\mu}^{a}\equiv e_{\mu}^{a}$ are identified as the spin-connection and the tetrade of the four-dimensional theory, respectively. Also, we introduce $\phi^{a}\equiv -l^{2}\Omega_{4}^{a4}$ and $\varphi\equiv lE_{4}^{4}$ as components of a scalar multiplet in four dimensions (usually called the radion); note that both $\phi^{a}$ and $\varphi$ have length dimension one. We make a standard truncation of the theory by keeping only four-dimensional graviton $(\omega_{\mu}^{ab}, e_{\mu}^{a})$ and four-dimensional scalar $(\phi^{a}, \varphi)$, and setting the components $\Omega_{4}^{ab}$, $E_{4}^{a}$, $\Omega_{\mu}^{a4}$, and $E_{\mu}^{4}$ to zero.

As for the curvature components, only $R_{\mu\nu}^{ab}$ and $R_{\mu 4}^{a4}$ remain after the truncation. In particular,
\begin{align}
R_{\mu\nu}^{ab}&=\partial_{\mu}\omega_{\nu}^{ab}-\partial_{\nu}\omega_{\mu}^{ab}+\omega_{\mu c}^{a}\omega_{\nu}^{cb}-\omega_{\nu c}^{a}\omega_{\mu}^{cb}, \nonumber\\
R_{\mu 4}^{a4}&=\partial_{\mu}\Omega_{4}^{a4}+\omega_{\mu c}^{a}\Omega_{4}^{c4}=-l^{-2}D_{\mu}\phi^{a},\nonumber\\
R_{\mu 4}^{ab}&=R_{\mu\nu}^{a4}=0.
\end{align} 
Thus we come to the reduced action in four dimensions, formulated in the coordinate-free fashion, 
\begin{align}\label{Sreduced}
S_{red}&=\frac{k(2\pi R)}{8l^{2}}\int\varepsilon_{abcd}
\Bigg(R^{ab}R^{cd}\varphi+\frac{2}{l^{2}}R^{ab}e^{c}e^{d}\varphi\\
+&\frac{1}{l^{4}}e^{a}e^{b}e^{c}e^{d}\varphi
-\frac{4}{3l^{3}}e^{a}e^{b}e^{c}D_{\omega}\phi^{d}
-\frac{4}{l}e^{a}R^{bc}D_{\omega}\phi^{d}
\Bigg).\nonumber
\end{align}
This action is invariant under $SO(3,2)$ gauge transformations, however this invariance is not manifest. Generators of the $SO(3,2)$ group are $J_{ab}$ and $J_{a5}$, with $SO(3,2)$ gauge field $1$-form and field strength $2$-form
\begin{align}
\mathcal{A}&=\frac{1}{2}\omega^{ab}J_{ab}+l^{-1}e^{a}J_{a5},\nonumber\\
\mathcal{F}&=\frac{1}{2}F^{ab}J_{ab}+F^{a5}J_{a5}.
\end{align}
By introducing adjoint scalar field  
\begin{align}
\Phi&=\Phi^{a}J_{4a}+\Phi^{5}J_{45}=\phi^{a}J_{4a}+\varphi J_{45},  
\end{align}
the reduced action (\ref{Sreduced}) becomes 
\begin{align}\label{FFP}
S_{red}&=\frac{ik(2\pi R)}{l^{2}}\int\tr\left(\mathcal{F}\mathcal{F}\Phi\right),
\end{align}
which is a topological action manifestly invariant under $SO(3,2)$ gauge transformations.

After the KK reduction and truncation, $SO(3,2)$ gauge transformations with parameter $\epsilon=\frac{1}{2}\epsilon^{ab}J_{ab}+\epsilon^{a5}J_{a5}$ (we set $\epsilon^{a4}=\epsilon^{45}=0$), induce the following variations of the gauge field components, 
\begin{align} 
\delta_{\epsilon}\omega^{ab}&=-\diff\epsilon^{ab}-\left(\omega^{a}_{\;\;c}\epsilon^{cb}+l^{-1}e^{a}\epsilon^{b5}-(a\leftrightarrow b)\right),
\nonumber\\
l^{-1}\delta_{\epsilon}e^{a}&=-\diff\epsilon^{a5}-\omega^{a}_{\;\;b}\epsilon^{b5}-l^{-1}e^{b}\epsilon_{b}^{\;\;a},\nonumber\\
\delta_{\epsilon}\phi^{a}&=\epsilon^{a}_{\;\;b}\phi^{b}+\epsilon^{a}_{\;\;5}\varphi,\nonumber\\
\delta_{\epsilon}\varphi&=-\epsilon^{a5}\phi_{a}.
\end{align}
with the remaining components being invariant. In terms of the $SO(3,2)$ connection $\mathcal{A}$ and the scalar field $\Phi$, these variations can be expressed as
\begin{align}
\delta_{\epsilon}\mathcal{A}&=-\diff\epsilon-[\epsilon,\mathcal{A}],\\
\delta_{\epsilon}\Phi&=[\epsilon,\Phi],
\end{align}
and therefore also $\delta_{\epsilon}\mathcal{F}=[\epsilon,\mathcal{F}]$. The $SO(3,2)$ invariance of the action (\ref{FFP}) now follows directly from the graded cyclicity of the trace.     

The local $SO(3,2)$ invariance of the theory can be reduced to the local Lorentz $SO(3,1)$ invariance by choosing $\phi^{a}=0$ and $\varphi=l$, which gives us
\begin{align}\label{klasicnoSB}
S_{red}=&\frac{k(2\pi R)}{8l^{3}}\int\varepsilon_{abcd}\nonumber\\
\times&\left(l^{2}R^{ab}R^{cd}
+2R^{ab}e^{c}e^{d}+\frac{1}{l^{2}}e^{a}e^{b}e^{c}e^{d}
\right).
\end{align}
The above action consists of the standard EH term, the cosmological constant term with $\Lambda=-3/l^{2}$, and the topological Gauss-Bonnet term that does not affect classical field equations; the four-dimensional gravitational constant is $G^{(4)}=\frac{l^{3}}{16\pi^{2}Rk}=\frac{G^{(5)}}{2\pi R}$.

Classical field equations are simply
\begin{align}\label{klasicnoe}
\varepsilon_{abcd}\left(R^{ab}+\frac{1}{l^2} e^a e^b \right)e^c&=0,\\
\varepsilon_{abcd}T^{a}e^{b}&=0.\label{notorsion}
\end{align}
Equation (\ref{notorsion}) shows that the classical geometry is torsionless, i.e. $T^{a}=0$, which is in agreement with the usual perception of Einstein's General Relativity. The first equation is the standard Einstein field equation with a negative cosmological constant.
The above analysis shows how KK reduction of the classical $D=5$ CS theory for the conformal gauge group $SO(4,2)$ leads to Einstein's gravity with negative cosmological constant in four dimensions. 
 
In what follows, we will study further this relation between $D=5$ CS theory and four-dimensional gravity in the setting of NC field theory.

%%%%%%%%%%%%%%%%%%%%%%%%%%%%%%%%%%%%%%%%%

\section{Noncommutative gauge field theory and the geometric Seiberg-Witten map}

To make a transition to NC gauge 
field theory, we will introduce some basic elements of the generalized $\star$-product formalism and the geometric version of the Seiberg-Witten (SW) map (a generalization of the ordinary SW map associated with the MWG $\star$-product \cite{SW}), closely following the account given in \cite{Leo}, where the geometric SW map was applied in the case of NC CS gravity. For completeness, we will recapitulate the main steps and results obtained in the case of NC $D=5$ CS gravity (they differ from those in \cite{Leo} only due to our slightly different conventions) in a manner that will be sufficient to proceed with the KK reduction of this theory in the following section. For the complete analysis and derivation of the NC correction to the classical CS gravity action in all odd dimensions, we point the reader to \cite{Leo}.       

NC gauge field theory (including NC gravity and supergravity) can be formulated in the language of twisted differential geometry, a short review of which can be found in \cite{PL_fer}. In the case of an abelian Drinfeld twist \cite{PLM-13, Aschieri:2011ng, Leo, Paolo, Castellani_SUGRA}, one introduces an associative exterior product between forms, denoted by $\wedge_{\star}$, that amounts to a deformation of the ordinary exterior product.  
For a $p$-form $\tau_{p}$ and a $q$-form $\tau'_{q}$, it is defined by 
\begin{widetext}
\begin{align}\label{NC_exterior}
\tau_{p}\wedge_{\star}\tau'_{q}&\equiv\sum\limits_{n=0}^{+\infty}\left(\frac{i}{2}\right)^{n}\theta^{I_{1}J_{1}}\dots\theta^{I_{n}J_{n}}(\ell_{I_{1}}\dots\ell_{I_{n}}\tau_{p})\wedge(\ell_{J_{1}}\dots\ell_{J_{n}}\tau_{q})  \nonumber\\
&=\tau_{p}\wedge\tau'_{q}+\frac{i}{2}\theta^{IJ}(\ell_{I}\tau_{p})\wedge(\ell_{J}\tau'_{q})+\frac{1}{2!}\left(\frac{i}{2}\right)^{2}\theta^{I_{1}J_{1}}\theta^{I_{2}J_{2}}(\ell_{I_{1}}\ell_{I_{2}}\tau_{p})\wedge(\ell_{J_{1}}\ell_{J_{2}}\tau'_{q})+\dots=\tau_{p}\wedge\tau'_{q}+\ell_{I}k^{I}.
\end{align}
\end{widetext}

where NC parameters $\theta^{IJ}$ $(I,J=1,\dots, s\leq D)$ comprise a constant antisymmetric matrix, while $\ell_{I}$ stand for Lie derivatives along mutually commuting vector fields $X_{I}$ (summation over indices $I,J$ is implied). The NC exterior product (\ref{NC_exterior}) is associative due to $[X_{I},X_{J}]=0$. The last line of (\ref{NC_exterior}) follows from the fact that, in the case of commuting vector fields, Lie derivatives along these fields also commute $[\ell_{I},\ell_{J}]=0$. Therefore, the $\wedge_{\star}$ product of forms differs form the ordinary exterior product by a total Lie derivative of some $(p+q)$-form $k^{I}$, which is given by the remaining summation in (\ref{NC_exterior}). 

A system of commuting vector fields $\{X_{I}\}$ that enter the definition of $\wedge_{\star}$ is an independent structure on the spacetime manifold that provides a coordinate-free formulation of an NC theory. However, we can choose (at least locally) a system of coordinates $x^{\mu}$ that is adapted to $\{X_{I}\}$ by setting $X_{I}=\partial/\partial x^{\mu}$ (note that the number of vector fields $X_{I}$ need not be equal to the number of spacetime dimensions $D$, in which case $\{X_{I}\}$ span a subspace of a $D$-dimensional tangent space), so that if $\tau_{p}$ and $\tau'_{q}$ are $0$-forms, the  (\ref{NC_exterior}) reduces to the MWG $\star$-product of functions (\ref{Moyal}) mentioned in the introduction. In particular, the $\star$-commutator between coordinate functions (treated as scalar fields) is simply $[x^{\mu},x^{\nu}]_{\star}=i\theta^{IJ}\delta^{\mu}_{I}\delta^{\nu}_{J}=\theta^{\mu\nu}=const$. In some other system of coordinates, unrelated to $\{X_{I}\}$, the $\star$-commutator would in general not be constant. In other words, the choice of $\{X_{I}\}$ determines in which coordinates do we have constant noncommutativity. At this point, it might seem artificial to introduce such a structure without giving it some physical interpretation. Although it is essential for defining a star-wedge product, it makes an NC theory emphatically not background independent. It is therefore natural to ask is there some principle that explains how these vector fields come about. 
Presumably, the criterion for choosing a particular set of commuting vector fields could come from some more fundamental theory that has NC field theory as its low-energy limit. 

Gauge field theory actions in the NC setting have the same form as the corresponding commutative ones, and are obtained directly from them by replacing ordinary exterior product with the NC-deformed exterior product $\wedge_{\star}$. However, to maintain gauge invariance of the NC theory, one has to introduces NC fields (we will denote them by a hat symbol) that change under NC gauge transformations in the same manner as classical fields change under ordinary gauge transformations. For an NC gauge parameter $\widehat{\epsilon}$, infinitesimal NC variations of NC gauge field $\widehat{A}$ and NC field strength $\widehat{F}$ are thus       
\begin{align}\label{NCvarA}
\widehat{\delta}_{\widehat{\epsilon}}\widehat{A}&=-\diff\widehat{\epsilon}-\widehat{A}\wedge_{\star}\widehat{\epsilon}+\widehat{\epsilon}\wedge_{\star}\widehat{A},\\
\widehat{\delta}_{\widehat{\epsilon}}\widehat{F}&=\widehat{\epsilon}\wedge_{\star}\widehat{F}-\widehat{F}\wedge_{\star}\widehat{\epsilon}. \label{NCvarF}
\end{align}
which amounts to an NC deformation of (\ref{varA}) and (\ref{varF}). The NC action is invariant under these deformed gauge transformations, by construction. 

In particular, the NC $D=5$ CS action reads
\begin{align}
S_{CS,NC}^{(5)}=-\frac{ik}{3}\int\tr\Big(&\widehat{F}\wedge_{\star}\widehat{F}\wedge_{\star}\widehat{A}\nonumber\\
-\frac{1}{2}&\widehat{F}\wedge_{\star}\widehat{A}\wedge_{\star}\widehat{A}\wedge_{\star}\widehat{A}\nonumber\\
+\frac{1}{10}&\widehat{A}\wedge_{\star}\widehat{A}\wedge_{\star}\widehat{A}\wedge_{\star}\widehat{A}\wedge_{\star}\widehat{A}
\Big).
\end{align}
A commutator between two infinitesimal NC gauge transformations acts on the (adjoint) field $F$ as
\begin{align}
[\widehat{\delta}_{\widehat{\epsilon}_{1}},\widehat{\delta}_{\widehat{\epsilon}_{2}}]\widehat{F}=\widehat{\delta}_{-[\widehat{\epsilon}_{1},\widehat{\epsilon}_{2}]_{\star}}\widehat{F}=-[[\widehat{\epsilon}_{1},\widehat{\epsilon}_{2}]_{\star},\widehat{F}]_{\star},    
\end{align}
with
\begin{equation}
[\widehat{\epsilon}_{1},\widehat{\epsilon}_{2}]_{\star}=\frac{1}{2}\left([\widehat{\epsilon}_{1}^{K},\widehat{\epsilon}_{2}^{L}]_{\star}\{T_{K},T_{L}\}+\{\widehat{\epsilon}_{1}^{K},\widehat{\epsilon}_{2}^{L}\}_{\star}[T_{K},T_{L}]\right).  
\end{equation}
The fact that anticommutator $\{T_{K},T_{L}\}$ appears in the above formula implies that infinitesimal NC gauge transformations are not generally closed in the Lie algebra. A way out of this difficulty is to consider a larger, universal enveloping algebra (UEA) of the original Lie algebra, i.e., to assume that the NC gauge transformations parameter is UEA-valued, implying that the NC gauge field $\widehat{A}$ and NC field strength $\widehat{F}$ are also UEA-valued. 

However, since UEA is infinite-dimensional, this leads to an infinite number of new degrees of freedom, which we find physically unacceptable. Seiberg-Witten (SW) map allows us to redefine these new NC degrees of freedom in terms of the original classical ones \cite{SW, UEA}. The main idea of Seiberg and Witten was to show that these UEA-valued NC fields that transform under NC gauge transformations can be organized into a perturbation series in powers of $\theta$, with coefficients built out of fields from the commutative theory ($\theta=0$) subjected to ordinary gauge transformation laws. 
The basic principle behind SW construction is that NC gauge transformations are induced by the corresponding commutative ones,
\begin{equation}
\widehat{\delta}_{\widehat{\epsilon}}\widehat{A}(A)=\widehat{A}(A+\delta_{\epsilon}A)-\widehat{A}(A),    
\end{equation}
where the NC gauge field $\widehat{A}$ is a function of the classical gauge field $A$ and $\widehat{\epsilon}=\widehat{\epsilon}(\epsilon,A)$. Using (\ref{varA}) and (\ref{NCvarA}) we can solve this differential equation perturbatively and derive the nonlinear SW map that represents NC fields $\widehat{A}$ and $\widehat{\epsilon}$ as a power series in $\theta$, with coefficients built out solely of classical fields. For an arbitrary set of commuting vector fields (for any abelian twist), SW expansions of the NC gauge field and NC parameter, up to first-order in $\theta$, are given by
\begin{align}
\hat{A}=A-\frac{i}{4}\theta^{IJ}\{A_{I},\ell_{J}A+F_{J}\},\\
\hat{\epsilon}=\epsilon-\frac{i}{4}\theta^{IJ}\{A_{I},\ell_{J}\epsilon\}.
\end{align}
where $0$-form $A_{I}\equiv i_{X_{I}}(A)=\frac{1}{2}\Omega^{AB}_{I}J_{AB}+l^{-1}E^{A}_{I}J_{A5}$ is a contraction of the gauge field along vector field $X_{I}$, and similar for $1$-form $F_{J}$.

SW map allows us to expend NC actions in powers of $\theta$ and ensures the invariance of the action under ordinary gauge transformations, at each order. The leading order term ($\theta=0$) is the classical action, and higher-order terms represent $\theta$-dependent NC corrections that can be interpreted as new effective interactions for classical fields. 
A detailed computation of an NC gauge variation of the NC CS $(2n-1)$-form can be found in \cite{Leo}. The result is 
\begin{align}
\delta_{\theta}&\widehat{Q}_{CS}^{(2n-1)}=\frac{i}{2}\delta\theta^{IJ}\\
&\times\int\tr\left(F\Diff F_{I}\sum\limits_{k=0}^{n-3}(k+1)F^{n-3-k}F_{J}F^{k}
\right).   \nonumber
\end{align}
We see that for $n=1$ and $n=2$ the first-order NC correction vanishes; it appears only for $n\geq3$. In particular, for $n=3$, i.e. for $D=5$, the variation of the CS Lagrangian is \begin{equation}
\delta_{\theta}L_{CS,NC}^{(5)}=\frac{k}{6}\delta\theta^{IJ}\tr\left(F\wedge\Diff F_{I}\wedge F_{J}\right),   
\end{equation}
where we have
\begin{align}
\Diff F_{I}&=\diff F_{I}+[A,F_{I}]\nonumber\\
&=\frac{1}{2}\left(\Diff_{\Omega}F_{I}^{AB}+\frac{1}{l^{2}}\left(E^{A}T_{I}^{B}-E^{B}T_{I}^{A}\right)\right)J_{AB}\nonumber\\
&\;\;\;\;\;\;\;\;\;\;+\frac{1}{l}\left(\Diff_{\Omega}T_{I}^{A}+F_{I}^{AB}E_{B}\right)J_{A5}.
\end{align}
After calculating traces and some algebra, we get
\begin{align}\label{8_terms}
S_{CS,\theta}^{(5)}&=\frac{k\theta^{IJ}}{12}\\
\times\int\bigg(&F^{AB}(F_{I})_{BC}(\Diff_{\Omega} F_{J})^{C}_{\;\;A}+\frac{1}{l^{2}}F^{AB}(F_{I})_{BC}(T_{J})^{C}E_{A} \nonumber\\ +&\frac{1}{l^{2}}F^{AB}(T_{I})_{B}(\Diff_{\Omega} T_{J})_{A}+\frac{2}{l^{2}}F^{AB}(T_{I})_{B}(F_{J})_{AC}E^{C}\nonumber\\
+&\frac{1}{l^{2}}T^{A}(T_{I})^{B}(\Diff_{\Omega} F_{J})_{BA}+\frac{1}{l^{2}}T^{A}
(\Diff_{\Omega}T_{I})^{B}(F_{J})_{BA} \nonumber\\  +&\frac{1}{l^{2}}T_{A}(F_{I})^{AB}(F_{J})_{BC}E^{C}+\frac{2}{l^{4}}T_{A}(T_{I})_{B}(T_{J})^{[B}E^{A]}
\bigg).\nonumber
\end{align}

This $\theta$-dependent action is manifestly
diffeomorphism-invariant and invariant under local $SO(4,2)$ gauge transformations by the general feature of the Seiberg-Witten construction. It represents the leading-order perturbative correction to the classical $D=5$ CS theory due to spacetime noncommutativity. 

%%%%%%%%%%%%%%%%%%%%%%%%%%%%%%%%%%%%%%%%

\section{Kaluza-Klein reduction of NC $D=5$ CS theory}

Now we come to the main step of our analysis, which is the KK reduction of the first-order NC action (\ref{8_terms}). Since this part is mainly technical, we only present the obtained results. However, to provide some insight into the details of the computational procedure, in Appendix C, we present an explicit example of the reduction of one of the terms in (\ref{8_terms}). 
We saw in Section 2 that KK reduction of the classical $D=5$ CS action with conformal gauge group $SO(4,2)$, followed by a suitable symmetry breaking necessary to obtain a Lorentz invariant theory, leads to the standard EH action with the negative cosmological constant in four dimensions. Now we apply the same procedure on $D=5$ first-order NC CS action (\ref{8_terms}) to obtain NC corrections to Einstein's gravity in four dimensions.  

First, we must see what happens to the basic building blocks under the KK reduction and symmetry breaking (SB).
The $F^{AB}$ components of the field strength $F=\tfrac{1}{2}F^{AB}J_{AB}+l^{-1}T^{A}J_{A5}$ after the KK reduction and SB are given by
\begin{align}
F_{\mu\nu}^{ab}&=R_{\mu\nu}^{ab}+\frac{1}{l^{2}}(e_{\mu}^{a}e_{\nu}^{b}-e_{\nu}^{a}e_{\mu}^{b}), \nonumber\\
F_{\mu 4}^{ab}&=F_{\mu\nu}^{a4}=0,\nonumber\\
F_{\mu 4}^{a4}&=-\frac{1}{l^{2}}D_{\mu}\phi^{a}+\frac{1}{l^{3}}e_{\mu}^{a}\varphi\xrightarrow{\text{SB}}\frac{1}{l^{2}}e_{\mu}^{a},
\end{align}
and torsion components are
\begin{align}
T_{\mu\nu}^{a}&=\partial_{\mu}e^{a}_{\nu}-\partial_{\nu}e^{a}_{\mu}+\omega^{a}_{\mu b}e_{\nu}^{b}-\omega^{a}_{\nu b}e_{\mu}^{b},\nonumber\\
T_{\mu 4}^a&=T_{\mu\nu}^4=0,\nonumber\\
T_{\mu 4}^4&=\frac{1}{l}D_\mu \varphi+\frac{1}{l^{2}}e_{\mu}^{a}\phi_a\xrightarrow{\text{SB}} 0.
\end{align}
Since they appears in (\ref{8_terms}), we also consider the covariant derivatives $(\Diff_{\Omega}F_{I})^{AB}$ and  $(\Diff_{\Omega}T_{I})^{A}$. The former is given by
\begin{align}
(\Diff_{\Omega}&F_{I})^{AB}=\diff F_{I}^{AB}+\Omega^{A}_{\;\;C}F_{I}^{CB}-\Omega^{B}_{\;\;C}F_{I}^{CA}\nonumber\\
=&\frac{1}{2}\Big[\partial_{\widetilde{\mu}}(X_{I}^{\widetilde{\alpha}}F_{\widetilde{\alpha}\widetilde{\nu}}^{AB})-\partial_{\widetilde{\nu}}(X_{I}^{\widetilde{\alpha}}F_{\widetilde{\alpha}\widetilde{\mu}}^{AB})\nonumber\\
&+X^{\widetilde{\alpha}}_{I}\Big(\Omega_{\widetilde{\mu}\;C}^{A}F_{\widetilde{\alpha}\widetilde{\nu}}^{CB}-\Omega_{\widetilde{\mu}\;C}^{B}F_{\widetilde{\alpha}\widetilde{\nu}}^{CA}\nonumber\\
&-\Omega_{\widetilde{\nu}\;C}^{A}F_{\widetilde{\alpha}\widetilde{\mu}}^{CB}+\Omega_{\widetilde{\nu}\;C}^{B}F_{\widetilde{\alpha}\widetilde{\mu}}^{CA}\Big)
\Big]\diff x^{\widetilde{\mu}}\diff x^{\widetilde{\nu}}, 
\end{align}
or, in components,
\begin{align}
(\Diff_{\Omega}F_I)^{ab}_{\mu\nu}&=D_{\mu}(X_{I}^{\alpha}F_{\alpha\nu}^{ab})-D_{\nu}(X_{I}^{\alpha}F_{\alpha\mu}^{ab}),\nonumber\\
(\Diff_{\Omega} F_I)^{a4}_{\mu\nu}&=-D_\mu(X_I^4 F_{\nu 4}^{a4})+D_\nu(X_I^4 F_{\mu 4}^{a4}),\nonumber\\
&\xrightarrow{\text{SB}}
-\frac{1}{l^{2}}\left(D_\mu(X_I^4 e_{\nu}^a)-D_\nu(X_I^4 e_\mu^a) \right),\nonumber\\
(\Diff_{\Omega} F_I)^{ab}_{\mu 4}&=\frac{1}{l^{2}}X_{I}^{4}\left(\phi^{a}F_{\mu 4}^{b4}-\phi^{b}F_{\mu 4}^{a4}
\right)\xrightarrow{\text{SB}} 0, \nonumber\\
(\Diff_{\Omega} F_I)^{a4}_{\mu 4}&=D_{\mu}(X_{I}^{\alpha}F_{\alpha 4}^{a4})-\frac{1}{l^{2}}X_{I}^{\alpha}F_{\alpha\mu}^{ab}\phi_{b}\nonumber\\ 
&\xrightarrow{\text{SB}}\frac{1}{l^2}D_\mu (X_I^\alpha e_\alpha^a),
\end{align}
and the latter, 
\begin{align}
(\Diff_{\Omega}&T_{I})^{A}=\diff T_{I}^{A}+\Omega^{A}_{\;\;B}T_{I}^{B}\nonumber\\
=&\frac{1}{2}\Big[\partial_{\widetilde{\mu}}\Big(X_{I}^{\widetilde{\alpha}}T_{\widetilde{\alpha}\widetilde{\nu}}^{A}\Big)-\partial_{\widetilde{\nu}}\Big(X_{I}^{\widetilde{\alpha}}T_{\widetilde{\alpha}\widetilde{\mu}}^{A}\Big)\nonumber\\
+&X^{\widetilde{\alpha}}_{I}\Big(\Omega_{\widetilde{\mu}\;B}^{A}T_{\widetilde{\alpha}\widetilde{\nu}}^{B}-\Omega_{\widetilde{\nu}\;B}^{A}T_{\widetilde{\alpha}\widetilde{\mu}}^{B}\Big)
\Big]\diff x^{\widetilde{\mu}}\diff x^{\widetilde{\nu}}, 
\end{align}
or, in components,
\begin{align}
(\Diff_{\Omega} T_I)^a_{\mu\nu}&=D_{\mu}(X_{I}^{\alpha}T_{\alpha\nu}^{a})-D_{\nu}(X_{I}^{\alpha}T_{\alpha\mu}^{a}),\nonumber\\
(\Diff_{\Omega} T_I)^4_{\mu\nu}&=D_{\mu}(X_{I}^{4}T_{4\nu}^{4})-D_{\nu}(X_{I}^{4}T_{4\mu}^{4})\xrightarrow{\text{SB}} 0,\nonumber\\
(\Diff_{\Omega}T_I)_{\mu 4}^a&=-X_{I}^{4}T_{\mu 4}^{4}\phi^{a}\xrightarrow{\text{SB}} 0,\nonumber\\
(\Diff_{\Omega}T_I)^4_{\mu 4}&=D_{\mu}(X_{I}^{\alpha}T_{\alpha 4}^{4})
-X_{I}^{\alpha}T_{\alpha \mu}^{a}\phi_{a}\xrightarrow{\text{SB}} 0.      
\end{align}

Since we are interested only in the four-dimensional theory, we may assume that $\partial_{\mu}X_{J}^{4}=0$. This will not break diffeomorphism invariance in four dimensions. In this case, the reduced NC action up to first-order reads
\begin{align}\label{S_red_NC}
S_{red,NC}&=S_{red}+\frac{(2\pi R) k}{12} \theta^{I4}\nonumber\\
\times\int\Bigg[&\frac{2}{l^{4}}R^{ab}T_{a}(e_{I})_{b}-\frac{4}{l^{4}}T^{a}(R_{I})_{ab}e^{b}\nonumber\\
+&\frac{2}{l^{4}}R^{ab}(T_{I})_{a}e_{b}+\frac{6}{l^{6}}T^{a}e_{a}(e_{I})^{b}e_{b}\Bigg],
\end{align}
where $S_{red}$ is the commutative action (\ref{klasicnoSB}) and $\theta^{I4}\equiv \theta^{IJ}X_{J}^{4}$ is constant. Action (\ref{S_red_NC}) describes the gravitational sector of the KK-reduced $D=5$ NC CS theory. It is manifestly invariant under $4$-diffeomorphisms and local $SO(3,1)$ Lorentz transformations, and it represents a modification od Einstein's gravity due to spacetime noncommutativity. Additional $\theta$-dependent NC terms can be interpreted as new gravitational interactions that capture some quantum gravity effects. 

By varying (\ref{S_red_NC}) with respect to the tetrade and the spin-connection, one obtains first-order NC field equations (note that the fact that classical geometry is torsionless greatly simplifies the result of the variation) that describe a gravitational system beyond classical Einstein's gravity,
\begin{widetext}
\begin{align}\label{NC_equation_e}
\delta e_{d}:&\;\;\; \varepsilon_{abc}^{\;\;\;\;\;d}\left(R^{ab}e^c+\frac{1}{l^2} e^a e^b e^c\right)-\frac{\theta^{I4}}{3l}\left[\left(R^{db}+\frac{3}{l^{2}}e^{d}e^{b}\right)(\Diff_{\omega} e_{I})_{b}-2(\Diff_{\omega} R_{I})^{db}e_{b}
\right]=0, \\ \label{NC_equation_omega}
\delta\omega_{ac}:&\;\;\;\varepsilon^{ac}_{\;\;\;bd}T^{b}e^{d}+\frac{\theta^{I4}}{3l}\left[\frac{1}{2}R^{ab}e^{c}(e_{I})_{b}-\frac{1}{2}R^{cb}e^{a}(e_{I})_{b}+(R_{I})^{ab}e^{c}e_{b}-(R_{I})^{cb}e^{a}e_{b}+\frac{3}{l^{2}}e^{a}e^{b}e^{c}(e_{I})_{b}
\right]=0.    
\end{align}
\end{widetext}

A point that we want to emphasize is the importance of some $X_{I}$ having a non-zero component in the compactified direction. This circumstance is necessary for having a nonvanishing first-order NC correction, and it suggests that 
noncommutativity between the compactified coordinate $\hat{x}^{4}$ and the remaining ones plays an essential role. In other words, if we were to demand that only non-compactified coordinates of the the four-dimensional theory have non-trivial commutation relations between each other, i.e.  $[\hat{x}^{\mu},\hat{x}^{\nu}]=i\theta^{\mu\nu}$ and $[\hat{x}^{\mu},\hat{x}^{4}]=0$, the first-order NC correction would be exactly zero. Moreover, this happens to be something quite general. Namely, it was proven in \cite{UlasSaka:2007ue} that the order in which we apply the KK reduction and the NC deformation is not important if we assume that only non-compactified coordinates fail to commute mutually. One could first take a classical action in $D=5$ (with one spatial dimension compactified), make a transition to NC theory and then apply the KK reduction, or instead make a KK reduction of the classical $D=5$ action first, and then make a transition to the NC version of the reduced theory. In either way, the result would be the same.
Therefore, an NC extra spatial dimension is essential for having non-trivial first-order NC effects in the KK-reduced theory. This fact, together with the considerations in \cite{MDVR-14, Us-16, UsLetter} proving that SW construction for four-dimensional gauge theory with Lagrangian (\ref{FFP}) has zero first-order correction, is in agreement with our conclusion. Therefore, we can safely assume that non-compactified coordinates mutually commute, and that NC effects are associated exclusively to the existence of noncommutative compactified spatial dimension, i.e. we can assume that $[\hat{x}^{\mu},\hat{x}^{\nu}]=0$ and $[\hat{x}^{\mu},\hat{x}^{4}]=i\theta^{\mu 4}$.

It is interesting to note that, by imposing  $[\hat{x}^\mu,\hat{x}^4]=i\theta^{\mu 4}$, i.e. noncommutativity between the compactified dimension and uncompactified ones, we imply the following form of the uncertainty relations
\begin{equation}\label{neodredjenost}   \Delta \hat{x}^\mu \Delta \hat{x}^4\geq \frac{|\theta^{\mu 4}|}{2}.
\end{equation}
On the other hand, as we compactified the fourth dimension on a circle of radius $R$, it makes sense to write $\Delta \hat{x}^4\sim R$. Therefore, $\Delta \hat{x}^\mu\geq\frac{|\theta^{\mu 4}|}{2R}$. From this relation follows that the NC length scale satisfies $l_{NC}=\sqrt{\theta}\sim R$, meaning that the NC structure of spacetime could have a deeper connection to the mechanism of compactification, the radius of compactification setting the scale of noncommutativity.

%%%%%%%%%%%%%%%%%%%%%%%%%%%%%%%%%%%%%%%%

\section{AdS-Schwarzschild solution and NC chiral gravitational anomaly}

We now analyze some solutions of the NC field equations (\ref{NC_equation_e}) and (\ref{NC_equation_omega}) obtained in the last section. At this point, we have to recall that our analysis is valid only perturbatively. This means that we should seek the solutions of our equations in the form of $e^a+\tilde{e}^a$ (and similar for the spin-connection), where $e^a$ stands for the commutative part of the tetrade that satisfies classical ($\theta^{I4}=0$) equations of motion (\ref{klasicnoe}) and (\ref{notorsion}), and $\tilde{e}^a, \tilde{\omega}^{ab}\sim \theta$ is the leading order NC correction satisfying the following first-order equations
\begin{widetext}
\begin{align}\label{epart}
\varepsilon_{abc}^{\;\;\;\;\;d}\left[\tilde{R}^{ab}e^c+\left(R^{ab}+\frac{3}{l^2} e^a e^b \right)\tilde{e}^c\right]=\frac{\theta^{I4}}{3l}\left[\left(R^{db}+\frac{3}{l^{2}}e^{d}e^{b}\right)(\Diff_{\omega} e_{I})_{b}-2(\Diff_{\omega} R_{I})^{db}e_{b}
\right],
\end{align}
\begin{align}\label{wpart}
\varepsilon^{ac}_{\;\;\;bd}\tilde{T}^{b}e^{d}=
-\frac{\theta^{I4}}{3l}\left[\frac{1}{2}R^{ab}e^{c}(e_{I})_{b}-\frac{1}{2}R^{cb}e^{a}(e_{I})_{b}+(R_{I})^{ab}e^{c}e_{b}-(R_{I})^{cb}e^{a}e_{b}+\frac{3}{l^{2}}e^{a}e^{b}e^{c}(e_{I})_{b}
\right].
\end{align}
\end{widetext}
Here, we have 
\begin{align}
\tilde{R}^{ab}&=\diff\tilde{\omega}^{ab}+\tilde{\omega}^{a}_{\;\;c}\omega^{cb}+\omega^{a}_{\;\;c}\tilde{\omega}^{cb},  \\
\tilde{T}^{a}&=\diff\tilde{e}^{a}+\tilde{\omega}^{a}_{\;\;b}e^{b}+\omega^{a}_{\;\;b}\tilde{e}^{b}.
\end{align}
In seeking a solution for (\ref{epart}) and (\ref{wpart}), we will restrict ourselves to the case of invertible tetrads.

It is well known that classical equation (\ref{klasicnoe}) is satisfied for maximally symmetric four-dimensional spacetime with negative curvature, i.e. the AdS spacetime. In a suitable coordinate system, the metric for this geometry is given by
\begin{equation}
\diff s^2=-\left(1+\frac{r^2}{l^2}\right)\diff t^2+\frac{1}{\left (1+\frac{r^2}{l^2}\right)}\diff r^2 + r^2\diff \Omega^2.
\end{equation}
This solution satisfies  $R^{ab}=-l^{-2}e^a e^b$. It is not hard to see that, due to this relation, the right-hand sides of both equations (\ref{epart}) and (\ref{wpart}) vanish, implying there is no first-order NC correction to the tetrade and the spin-connection for AdS spacetime; it remains a valid solution even in NC theory, at least to first order in $\theta$.

Another interesting solution to consider is the AdS-Schwarzschild black hole, with metric
\begin{equation}\label{AdSBH}
    \diff s^2=-f^2(r)\diff t^2+\frac{1}{f^2(r)}\diff r^2 + r^2\diff \Omega^2,
\end{equation}
with $f^2(r)\equiv \left(1-\frac{2m}{r}+\frac{r^2}{l^2}\right) $. The horizon's position is determined by $f^2(r)=0$. A simple analysis shows that there is only one real solution to this equation, and therefore only one horizon, see for example \cite{Socolovsky:2017nff}. This metric, which includes a black hole,
a white hole, and two asymptotically AdS causally disconnected
spacetimes is an extraordinary laboratory to study black
hole physics in a non asymptotically flat spacetime \cite{Hawking:1982dh}. In particular,
the Schwarzschild case offers the possibility to discuss singularities, horizons
and boundaries in a simple but non-trivial way. 
We can now choose tetrads as 
\begin{align}
    e^0=f(r)\diff t,\hspace{4mm}
    e^1=\frac{\diff r}{f(r)},\hspace{4mm}
    e^2=r\diff \theta,\hspace{4mm}
    e^3=r\sin \theta\diff \phi.
\end{align}
As torsion vanishes, we have $T^{a}=\diff e^a+\omega^{ab}e_b=0$, and therefore nonzero components of the spin-connnection are 
\begin{align}\nonumber
    \omega^{01}&=\frac{1}{2}(f^2(r))'\diff t,\hspace{6mm}
    \omega ^{12}=-f(r)\diff\theta, \\
    \omega^{13}&=-f(r)\sin\theta\diff\phi, \hspace{3.5mm}
    \omega^{23}=-\cos\theta\diff\phi.
\end{align}
Components of the curvature tensor are then
\begin{align}\nonumber
&R^{01}=-\frac{1}{2}(f^{2}(r))''e^{0}e^{1},\hspace{2mm}
R^{02}=-\frac{1}{2r}(f^{2}(r))'e^{0}e^{2},\\\nonumber
&R^{03}=-\frac{1}{2r}(f^{2}(r))'e^{0}e^{3},\hspace{2mm}
R^{12}=-\frac{1}{2r}(f^{2}(r))'e^{1}e^{2},\\
&R^{13}=-\frac{1}{2r}(f^{2}(r))'e^{1}e^{3},\hspace{2mm}
R^{23}=\frac{1}{r^2}(1-f^{2}(r))e^{2}e^{3}.
\end{align}
We note that $R^{ab}$ has the general form $R^{ab}=F^{(a,b)}(r)e^{a}e^{b}$,  This will be important when solving for the NC corrections. 

We now propose an ansatz for the sought NC solution. Perhaps the most natural choice would be to assume that $e^0$ and $e^1$ get corrected by some function of $r$. However, NC field equations predict that tetrads do not acquire a first-order NC correction for a wide range of cases. 
There is, however, an even simpler option. We can assume that tetrads remain the same, and that only spin-connection acquires a first-order NC correction. Therefore, assuming $\tilde{e}^{a}=0$ and $\tilde{\omega}^{ab}\neq 0$, we have
\begin{align}\nonumber
\varepsilon_{abc}^{\;\;\;\;\;d}\tilde{R}^{ab}e^{c}=\frac{\theta^{I4}}{l}&\Bigg[\left(F^{(d,b)}(r)+\frac{1}{l^{2}}\right)(\Diff_{\omega} e_{I})_{b}\\
&+\frac{2}{3}\diff F^{(d,b)}(r)(e_{I})_{b}
\Bigg]e^{d}e^{b},
\end{align}
together with
\begin{align}\label{zaT}
\varepsilon^{ac}_{\;\;\;bd}\tilde{T}^{b}e^{d}=
\frac{\theta^{I4}}{2l}\left[F^{(a,b)}(r)+F^{(c,b)}(r)+\frac{2}{l^{2}}\right]e^{a}e^{c}e^{b}(e_{I})_{b}.
\end{align}
Equation (\ref{zaT}) can then be used to obtain all nonzero components of the NC torsion $\tilde{T}^{a}$. Until now, we did not assume any particular properties of the vector fields $\{X_I\}$, apart from the fact that $\partial_4 X_I ^{\widetilde{\mu}}=0$  and $\partial_\mu X_I^4=0$. We would like, however, to simplify our
equations. 

First, we will assume that non-compactified coordinates mutually commute, $[\hat{x}^\mu,\hat{x}^\nu]=0$, since this is not relevant at first-order in $\theta$. Furthermore, when dealing with quantum fields on a NC spacetime, non-vanishing value of $\theta^{0\widetilde{\mu}}$ components leads to a non-unitary evolution \cite{Buric:2005xe}; we avoid this issue by taking $[\hat{x}^0,\hat{x}^4]=0$.
Since the main idea is that NC effects are essentially associated with the existence of a compactified extra dimension, we will choose only two vector fields: $X_1=\partial_r$ and $X_4=\partial_4$. This choice leads to the $\star$-commutator $[r,x^4]_\star=i\theta^{14}$ which is directly related to the commutation relations $[\hat{r},\hat{x}^4]=i\theta^{14}$ in the abstract algebraic setting. In general, we are not aware of any definite rule of choosing the coordinates for which we impose the constant noncommutativity. One approach would be to consider vector fields generated by Killing vector fields of a commutative solution, which often simplifies the equations \cite{Aschieri:2011ih}. However, we have not decided to pursue this here, though it can be shown that most conclusions we obtain can be derived in the case where we choose only to include $X_3=\partial_\phi$. There are also different proposals in the literature \cite{UsLetter}. In any case, by identifying vector fields $X_I$ with basis vectors of some coordinate system, we manifestly break the diffeomorphism-invariance of a theory (it is like choosing a gauge). However, transformation rules for the vector fields enable us to find noncommutativity relations in any system of coordinates. 

Non-zero components of the NC torsion are 
\begin{align}
\tilde{T}_{23}^{0}&=-\frac{m\theta^{14}}{l}\frac{\sin\theta}{rf(r)},\nonumber\\
\tilde{T}_{03}^{2}&=\frac{m\theta^{14}}{2l}\frac{\sin\theta}{r^{2}},\nonumber\\
\tilde{T}_{02}^{3}&=-\frac{m\theta^{14}}{2lr^{2}}.
\end{align}
Next, we use $\tilde{T}^{a}=\diff\tilde{e}^{a}+\tilde{\omega}^{a}_{\;\;b}e^{b}+\omega^{a}_{\;\;b}\tilde{e}^{b}=\tilde{\omega}^{a}_{\;\;b}e^{b}$ to get the NC corrections for the spin-connection
\begin{align}\label{prvatorzija}
\tilde{\omega}_{0}^{23}&=\frac{m\theta^{14}}{lr^{3}}, \nonumber\\ 
\tilde{\omega}_{2}^{03}&=-\frac{m\theta^{14}}{2lr^{2}f(r)},\nonumber\\ 
\tilde{\omega}_{3}^{02}&=\frac{m\theta^{14}}{2lr^{2}f(r)}\sin\theta.
\end{align}
Finally, we can compute NC corrections to $R^{ab}$,
\begin{align}
\tilde{R}_{23}^{01}&=-\frac{m\theta^{14}}{l}\frac{\sin\theta}{r^{2}},\nonumber\\
\tilde{R}_{13}^{02}&=-\frac{m\theta^{14}}{l}\left[\frac{f'(r)}{2r^{2}}+\frac{f(r)}{r^{3}}\right]\frac{\sin\theta}{f^{2}(r)}=-\sin\theta\tilde{R}_{12}^{03},\nonumber\\
\tilde{R}_{03}^{12}&=\frac{m\theta^{14}}{l}\left[\frac{f'(r)}{2r^{2}}-\frac{f(r)}{r^{3}}\right]\sin\theta=-\sin\theta\tilde{R}_{02}^{13},\nonumber\\
\tilde{R}_{01}^{23}&=\frac{3m\theta^{14}}{lr^{4}}.
\end{align}
Having obtained a particular NC solution, we can ask how much it differs from its classical counterpart. For that matter, we will analyze $4$-forms whose integral over a compact manifold leads to certain topological invariants. For a general discussion see, for example, \cite{CS_book}. First consider the Nieh-Yan form, \begin{equation}
\mathcal{N}=T^aT_a-R^{ab}e_a e_b.
\end{equation}
Classically, this form vanishes for AdS-Schwarzschild, as the geometry is torsionless, and it is easy to see that it also vanishes when the first-order NC the corrections are included.

Next, we consider the Euler form
\begin{equation}
\mathcal{E}=\varepsilon^{abcd}R_{ab}R_{ed}.
\end{equation}
The classical computation gives us  
\begin{equation}
    \mathcal{E}=\left (\frac{48m^2}{r^6}+\frac{24}{l^4}    \right)r^2\sin \theta \;\diff t\wedge \diff r \wedge \diff \theta \wedge \diff \phi.
\end{equation}
However, a simple calculation shows that there is no correction to this result to first-order in $\theta^{14}$.

Finally, we consider the Pontryagin invariant. Pontryagin $4$-form is $\mathcal{P}=R^{ab}R_{ab}$. Classically, this form for the AdS-Schwarzschild spacetime is zero. However, there are non-trivial first-order NC contributions. We can easily check that 
\begin{align}
R^{ab}R_{ab}= \frac{48m^2\theta^{14}}{lr^5}\sin \theta \hspace{0.5mm}\diff t\wedge\diff r\wedge\diff \theta\wedge \diff \phi. \label{pontrjagin}
\end{align}
Note that this result is finite at the horizon, even though the coordinates we used are not defined there, and components of spin-connection are singular.
This result is interesting because Pontryagin density is directly related to the chiral gravitational anomaly. Let us briefly consider this anomaly here. 

Suppose we have a massless Dirac fermion coupled with some fixed geometric background. Kinetic term for this theory would be  \begin{equation}\label{kineticki}
    \sqrt{-g}\;\overline{\psi}\gamma^\mu\nabla_\mu \psi.
\end{equation} 
In massless electrodynamics, it is well-known that apart from the usual $U(1)$ charge conserving symmetry, the classical theory is invariant under global transformations $\psi\rightarrow e^{i\alpha\gamma_5}\psi$. Those transformations lead to separate conservation of the number of right-handed and left-handed fermions. However, this symmetry is anomalous, and upon quantization, there is no conservation of this type. The divergence of a classically conserved current is proportional to $F^2$, where $F$ is the curvature of the gauge connection \cite{Bertlmann}. We then expect that when coupled to gravity, the anomaly would be proportional to $R^{ab}R_{ab}$. This is further confirmed by a direct computation, which relies on the fact that the geometry is torsionless. When torsion is included, it is not a priori clear whether the anomaly will have additional terms. In \cite{Chandia:1997hu}, it was shown that the anomaly is given by a combination $\mathcal{P}+2\mathcal{N}$, suitably normalized. The derivation within, however, reveals that it is necessary to rescale the tetrade by a factor of a cut-off for $\mathcal{N}$ contribution to be finite. It was later argued in \cite{Kreimer:1999yp} that this rescaling was done inconsistently and that the final result should still be given in terms of Pontryagin density. However, in our case, $\mathcal{N}=0$, and as there are no other torsional invariants in four dimensions, we can safely use that the anomaly is given solely in terms of the Pontryagin density. More precisely, we have \cite{Bertlmann} 
\begin{equation}\label{petdvadeset}
    \diff * j_5=\frac{1}{96\pi^2}R^{ab}R_{ab},
\end{equation}
where $j_5$ is axial fermionic current, and $*$ is the Hodge dual operation with respect to the metric $g_{\mu\nu}=e_\mu^a e_{\nu a}$. Plugging in the expression (\ref{pontrjagin}), we obtain
\begin{equation}
    \diff * j_5=\frac{m^2\theta^{14}}{2\pi^2lr^5}\sin \theta \;\diff t \wedge \diff r \wedge \diff \theta\wedge \diff \phi,
\end{equation}
which is equivalent to 
\begin{equation}
    \partial_\mu (\sqrt{-g}j_5^\mu)=\frac{m^2\theta^{14}}{2\pi^2lr^5}\sin \theta.
\end{equation}
This expression is valid for all $r>r_h$ including large distances and small cosmological constant. An important thing to note is that this anomaly contribution vanishes in the strict limit of zero cosmological constant. Observational data from our universe suggest that the sign of the cosmological constant is positive, but we hope that the derived result can lead to a better understanding of anomalies induced by spacetime noncommutativity in more realistic physical settings. 

\section{Discussion}

The primary purpose of this paper was to investigate some aspects of the NC $D=5$ CS theory based on the AdS gauge group $SO(4,2)$, regarding its connection to the observable four-dimensional physics. The standard procedure of KK dimensional reduction led us to a model of four-dimensional gravity that amounts to a modification of Einstein's General Relativity with negative cosmological constant by $\theta$-dependent NC corrections as new effective interactions that ought to capture some quantum gravity effects. An important observation was that non-trivial NC contribution at first-order in $\theta$ comes only from noncommutativity between the compactified extra spatial dimension of the original CS theory and the remaining non-compactified spacetime dimensions. Field equations are derived from the effective four-dimensional NC gravity action, and, as an example, we studied how noncommutativity changes the torsionless classical geometry of the AdS-Schwarzschild black hole. Besides the modification of curvature and torsion, we found that Pontryagin density no longer vanishes, signalling that the NC-deformed AdS-Schwarzschild background gives rise to a chiral gravitational anomaly.                

We want to make several remarks, the first of which concerns the origin of the anomaly. We considered coupling massless Dirac fermions to a fixed NC AdS-Schwarzschild background geometry. One could object that the only consistent way to couple fermions would be to consider an NC theory of gravity together with fermions. This was done in the context of NC $SO(3,2)_\star$ gravity, and one can check that there is a non-trivial first-order NC correction to the classical Dirac action \cite{Gocanin:2017lxl}, see also \cite{Paolo}. However, we are only considering a quantized theory of massless Dirac fermions in a particular fixed NC background geometry. Apart from this general issue, one could also analyze other NC backgrounds, which would require a better understanding of the NC field equations and their solutions. The proposed setup could also analyze NC effects on black hole thermodynamics.    

We put fermions by hand in the four-dimensional theory in our anomaly computation, which is
another aspect that deserves further attention. Matter fields are typically present upon compactifying a pure gravity theory. 
In our analysis, we considered only zero-modes of the KK-expansion and focused only on the gravitational sector of the reduced theory. However, even if we decided to keep track of the KK matter fields (including the higher-modes), only bosons would be present in the compactified theory - another reason for neglecting matter fields altogether in the KK procedure, as fermions are missing. Regarding this issue, supersymmetric extensions of CS gravity Lagrangians are well-known \cite{CS}, and one might expect that the analysis from this paper, suitably modified to incorporate supergravity, would produce a more realistic theory as far as the field content is concerned. 

Another significant point is that we had to keep $R$ finite, which can be seen from the equation (\ref{neodredjenost}), as well as from the fact that it would make no sense to define pure four-dimensional theory with nonvanishing $\theta^{14}$ component. One may object that we should keep track of higher KK modes. However, our interest in this work lies solely in the gravitational part of the theory. We, therefore, insist on the ``vacuum'' solutions, those solutions whose energy-momentum tensor (and its spin counterpart) vanishes.

Also, in a braneworld scenario, matter fields are confined to a $D=4$ spacetime \cite{Arkani-Hamed}, and therefore are unaffected by noncommutativity driven by the NC parameter $\theta^{14}$. On the other hand, gravity is free to propagate in extra dimensions, and therefore gravity action develops NC corrections. Of course, based on the present results, we can not directly support this claim.

As for the appearance of the chiral gravitational anomaly, it would be interesting to check whether a different result would be obtained if one considers the full NC theory of massless fermions by including also heavy KK modes. We expect that even in this setting, there is a nonzero contribution to a chiral gravitational anomaly for the AdS-Schwarzschild solution. We base our expectation on the fact that we have explicitly computed the ''topological'' forms in the last section, and see no way of cancelling their (hypothetical) contributions; we also do not see an obvious modification due to noncommutativity that could cancel the obtained result. We postpone a full consideration of this issue for future work.  
Finally, there is a general issue concerning the interpretation of the vector fields used to define the NC version of the exterior product. Apart from allowing us to transition from one coordinate system to another covariantly, the physical criteria for selecting these fields is lacking. The most natural thing is to adapt them to some particular coordinate system, as we did at the end. However, one could also use more general vector fields to generalize the obtained results.
For example, one could insist not on taking $X_1=\partial_r$, but rather $X_1=g(r)\partial_r$. Motivation for this comes from the fact that many different coordinate systems can be related to the one we used in this paper by a suitable rescaling of the radial coordinate. For example, isotropic coordinates are more natural when considering gravity effects far away from a localized object. Radial coordinate $\rho$ in those coordinates is connected to $r$ precisely with the type of transformation discussed here. There is also a proposal to treat these vector fields as dynamical fields \cite{X_dynamical}.

{\bf Statements and Declarations}\newline
D.G. acknowledges the funding provided by the Faculty of Physics, University of Belgrade, through the grant by the Ministry of Education, Science, and  Technological Development of the Republic of Serbia  (451-03-68/2022-
14/200162). We thank Branislav Cvetkovi\'{c} for pointing out that computation of topological invariants could be of interest, and to Voja Radovanovi\'{c} and Marija Dimitrijevi\'{c} \'{C}iri\'{c} for useful discussions.

%%%%%%%%%%%%%%%%%%%%%%%%%%%%%%%%%%%%%%%%%%%%%%%%%%%%%%%%%%%%%%%

\appendix
\setcounter{figure}{0}
\renewcommand{\thefigure}{A\arabic{figure}}
\renewcommand{\theequation}{\Alph{section}.\arabic{equation}}
\section*{APPENDIX}

\section{Algebra of Lie algebra-valued forms}

For a Lie algebra $\mathfrak{g}$ with generators $\{T_{A}\}$, let $\alpha_{p}$ be a $\mathfrak{g}$-valued $p$-form and $\beta_{q}$ a $\mathfrak{g}$-valued $q$-form, i.e.
\begin{align}
\alpha_{p}&=\frac{1}{p!}\alpha_{\mu_{1}\dots\mu_{p}}\diff x^{\mu_{1}}\dots\diff x^{\mu_{p}}, \nonumber\\
\beta_{q}&=\frac{1}{q!}\beta_{\nu_{1}\dots\nu_{q}}\diff x^{\nu_{1}}\dots\diff x^{\nu_{q}},
\end{align}
where $\alpha_{\mu_{1}\dots\mu_{p}}=\alpha_{\mu_{1}\dots\mu_{p}}^{A}T_{A}$ and $\beta_{\nu_{1}\dots\nu_{q}}=\beta_{\nu_{1}\dots\nu_{q}}^{A}T_{A}$. 

Trace and (anti)-commutator are defined as
\begin{align}
\tr\left(\alpha_{p}\beta_{q}\right)&=\frac{1}{p!q!}\tr\left(\alpha_{\mu_{1}\dots\mu_{p}}\beta_{\nu_{1}\dots\nu_{q}}\right)\diff x^{\mu_{1}}\dots\diff x^{\nu_{q}},\nonumber\\
[\alpha_{p},\beta_{q}]&=\frac{1}{p!q!}[\alpha_{\mu_{1}\dots\mu_{p}},\beta_{\nu_{1}\dots\nu_{q}}]\diff x^{\mu_{1}}\dots\diff x^{\nu_{q}}, \nonumber\\
\{\alpha_{p},\beta_{q}\}&=\frac{1}{p!q!}\{\alpha_{\mu_{1}\dots\mu_{p}},\beta_{\nu_{1}\dots\nu_{q}}\}\diff x^{\mu_{1}}\dots\diff x^{\nu_{q}},
\end{align}
and the following ``graded'' properties hold:
\begin{align}
\tr\left(\alpha_{p}\beta_{q}\right)&=(-1)^{pq}\tr\left(\beta_{q}\alpha_{p}\right),\label{tr_graded}\\
[\alpha_{p},\beta_{q}]&=\alpha_{p}\beta_{q}-(-1)^{pq}\beta_{q}\alpha_{p}=-(-1)^{pq}[\beta_{q},\alpha_{p}], \nonumber\\
\{\alpha_{p},\beta_{q}\}&=\alpha_{p}\beta_{q}+(-1)^{pq}\beta_{q}\alpha_{p}=(-1)^{pq}\{\beta_{q},\alpha_{p}\}.\nonumber
\end{align}

Gauge invariance of $\tr\left(F^{n}\right)$ now follows from the graded cyclicity of the trace (\ref{tr_graded}),
\begin{align}\delta_{\epsilon}\tr\left(F^{n}\right)&=\tr\left(\delta_{\epsilon}F F^{n-1}+\dots+F^{n-1}\delta_{\epsilon}F\right)\nonumber\\&=n\tr\left([\epsilon,F] F^{n-1}\right)\nonumber\\
&=n\tr\left(\epsilon F^{n}-F\epsilon F^{n-1}\right)\nonumber\\&=0.
\end{align}

\section{Trace identities for $\Gamma$-matrices and $SO(4,2)$ generators}

Some relevant traces of $D=5$ $\Gamma$-matrices satisfying Clifford algebra $\{\Gamma_{A},\Gamma_{B}\}=2G_{AB}\mathbb{1}_{4\times 4}$, with metric $G_{AB}=(-++++)$, are
\begin{align}
&\tr(\Gamma_{A})=0,\nonumber\\
&\tr(\Gamma_{A}\Gamma_{B})=4G_{AB},\nonumber\\
&\tr(\Gamma_{A}\Gamma_{B}\Gamma_{C})=0,\nonumber\\
&\tr(\Gamma_{A}\Gamma_{B}\Gamma_{C}\Gamma_{D})=4(G_{AB}G_{CD}-G_{AC}G_{BD}+G_{BC}G_{AD}),\nonumber\\
&\tr(\Gamma_{A}\Gamma_{B}\Gamma_{C}\Gamma_{D}\Gamma_{E})=4i\varepsilon_{ABCDE},
\end{align}
and for $SO(4,2)$ generators $J_{AB}=\frac{1}{2}\Gamma_{AB}=\frac{1}{4}[\Gamma_{A},\Gamma_{B}]$ and $J_{A5}=\frac{1}{2}\Gamma_{A}$, we have
\begin{align}
\tr(J_{AB}J_{CD}J_{E5})&=\tfrac{i}{2}\varepsilon_{ABCDE},\nonumber\\
\tr(J_{AB}J_{C5}J_{D5})&=\tfrac{1}{2}(G_{AB}G_{CD}-G_{AC}G_{BD}+G_{BC}G_{AD}),\nonumber\\
\tr(J_{A5}J_{C5}J_{D5})&=0.
\end{align}

We define $\varepsilon_{01234}=+1$ and $\varepsilon^{01234}=-1$, with the following contractions: 
\begin{align}
\varepsilon^{ABCDE}\varepsilon_{ABCMN}&=-3!(\delta^{D}_{M}\delta^{E}_{N}-\delta^{D}_{N}\delta^{E}_{M}),\nonumber\\
\varepsilon^{ABCDE}\varepsilon_{ABCDF}&=-4!\delta^{E}_{F},\nonumber\\
\varepsilon^{ABCDE}\varepsilon_{ABCDE}&=-5!.
\end{align}

Denoting five-dimensional spacetime indices by $\widetilde{\mu}=(\mu,4)$, we have
\begin{equation}
\diff x^{\widetilde{\mu}}\wedge\diff x^{\widetilde{\nu}}\wedge\diff x^{\widetilde{\rho}}\wedge\diff x^{\widetilde{\sigma}}\wedge \diff x^{\widetilde{\tau}}=-\varepsilon^{\widetilde{\mu}\widetilde{\nu}\widetilde{\rho}\widetilde{\sigma}\widetilde{\tau}}\diff x^{5}.    
\end{equation}

\section{KK reduction and symmetry breaking}

To provide some insight into the details of the computation, we will focus on the fourth term in (\ref{8_terms}), i.e.
\begin{equation}
\frac{2\theta^{IJ}}{l^{2}}F^{AB}(T_{I})_{B}(F_{J})_{AC}E^{C},
\end{equation}
because it gives us, along with the first one and the seventh one, a nonvanishing contribution to the NC correction after the KK reduction and symmetry breaking; the remaining terms in (\ref{8_terms}) vanish. 

The first step is to ``unpack'' the differential forms and integrate out the extra coordinate $x^{4}$ (this gives us a factor of $2\pi R$ that we omit in the following formulae),
\begin{align}\label{4th_term}
-\frac{\theta^{IJ}}{l^{2}}X_{I}^{\widetilde{\alpha}}&X_{J}^{\widetilde{\beta}}F_{\widetilde{\mu}\widetilde{\nu}}^{AB}T_{\widetilde{\alpha}\widetilde{\rho}B}F_{\widetilde{\beta}\widetilde{\sigma}AC}E^{C}_{\widetilde{\tau}}\varepsilon^{\widetilde{\mu}\widetilde{\nu}\widetilde{\rho}\widetilde{\sigma}\widetilde{\tau}}\diff^{4}x  \\
=-\frac{\theta^{IJ}}{l^{2}}\Big(&X_{I}^{\alpha}X_{J}^{\beta}F_{\widetilde{\mu}\widetilde{\nu}}^{AB}T_{\alpha\widetilde{\rho}B}F_{\beta\widetilde{\sigma}AC}E^{C}_{\widetilde{\tau}}\nonumber\\ +&X_{I}^{4}X_{J}^{4}F_{\widetilde{\mu}\widetilde{\nu}}^{AB}T_{4\widetilde{\rho}B}F_{4\widetilde{\sigma}AC}E^{C}_{\widetilde{\tau}}\nonumber \\
+&X_{I}^{\alpha}X_{J}^{4}F_{\widetilde{\mu}\widetilde{\nu}}^{AB}T_{\alpha\widetilde{\rho}B}F_{4\widetilde{\sigma}AC}E^{C}_{\widetilde{\tau}}
\nonumber\\
+&X_{I}^{4}X_{J}^{\beta}F_{\widetilde{\mu}\widetilde{\nu}}^{AB}T_{4\widetilde{\rho}B}F_{\beta\widetilde{\sigma}AC}E^{C}_{\widetilde{\tau}}\Big)\varepsilon^{\widetilde{\mu}\widetilde{\nu}\widetilde{\rho}\widetilde{\sigma}\widetilde{\tau}}\diff^{4}x. \nonumber
\end{align}

The four terms in the last expression come from the fact that, in general, vector fields $\{X_{I}\}$ have five components $X_{I}^{\widetilde{\alpha}}=(X_{I}^{\alpha},X_{I}^{4})$ with $\alpha=0,1,2,3$. We immediately see that the second term vanishes due to $\theta^{IJ}X_{I}^{4}X_{J}^{4}=0$. Moreover, in turns out that, after the KK reduction, the $X_{I}^{\alpha}X_{J}^{\beta}$ term also vanishes. This result is in accordance with \cite{UlasSaka:2007ue}. Namely, if we were to assume that only non-compactified coordinates fail to commute, i.e. $[\hat{x}^{\mu},\hat{x}^{\nu}]=i\theta^{\mu\nu}$ while $[\hat{x}^{\mu},\hat{x}^{4}]=0$, there would be no NC correction at first-order in $\theta$, and we would have to compute second-order perturbations, which is much more difficult. Therefore, the expression (\ref{4th_term}) comes down to the two ``mixed'' terms, which can be combined into a single term,
\begin{align}
-\frac{\theta^{IJ}}{l^{2}}X_{I}^{\alpha}X_{J}^{4}\Big(&F_{\widetilde{\mu}\widetilde{\nu}}^{AB}T_{\alpha\widetilde{\rho}B}F_{4\widetilde{\sigma}AC}E^{C}_{\widetilde{\tau}}\nonumber\\
-&F_{\widetilde{\mu}\widetilde{\nu}}^{AB}T_{4\widetilde{\rho}B}F_{\alpha\widetilde{\sigma}AC}E^{C}_{\widetilde{\tau}}\Big)\varepsilon^{\widetilde{\mu}\widetilde{\nu}\widetilde{\rho}\widetilde{\sigma}\widetilde{\tau}}\diff^{4}x \nonumber\\
=\frac{\theta^{IJ}}{l^{2}}X_{I}^{\alpha}X_{J}^{4}\Big(-&F_{\mu\nu}^{AB}T_{\alpha\rho B}F_{4\sigma AC}E^{C}_{4}
+2F_{\mu 4}^{AB}T_{\alpha\nu B}F_{4\rho AC}E^{C}_{\sigma}\nonumber\\
-&2F_{\mu 4}^{AB}T_{4\nu B}F_{\alpha\rho AC}E^{C}_{\sigma}\Big)\varepsilon^{\mu\nu\rho\sigma}\diff^{4}x \nonumber \\
=\frac{\theta^{IJ}}{l^{2}}X_{I}^{\alpha}X_{J}^{4}\Big(&F_{\mu\nu}^{ab}T_{\alpha\rho b}F_{\sigma 4 a4}E^{4}_{4}
-2F_{\mu 4}^{b4}T_{\alpha\nu b}F_{\rho 4 c4}E^{c}_{\sigma}\nonumber\\
+&2F_{\mu 4}^{a4}T_{\nu 4}^{ 4}F_{\alpha\rho ac}E^{c}_{\sigma}\Big)\varepsilon^{\mu\nu\rho\sigma}\diff^{4}x \nonumber \\
\xrightarrow{\text{SB}} -\frac{\theta^{IJ}}{l^{4}}&X_{I}^{\alpha}X_{J}^{4}R_{\mu\nu}^{ab}T_{\alpha\rho a}e_{\sigma b}\varepsilon^{\mu\nu\rho\sigma}\diff^{4}x.
\end{align}
The last line is the result after the symmetry breaking.

\end{document}